\documentclass[pre,superscriptaddress,twocolumn,showpacs]{revtex4}
\usepackage{graphicx,bbm}
\usepackage{hyperref}
\usepackage{amssymb,amsfonts,amsmath}
\usepackage{float}
\def\etal#1{ {\em et al.}}
\def\tit#1{}
\def\ii{{\rm i}}

\def\llan{\left\langle}
\def\rran{\right\rangle}
\makeatletter
\def\underbracket{\@ifnextchar [ {\@underbracket} {\@underbracket [\@bracketheight]}}
\def\@underbracket[#1]{\@ifnextchar [ {\@under@bracket[#1]} {\@under@bracket[#1][0.4em]}}
\def\@under@bracket[#1][#2]#3{%\message {Underbracket: #1,#2,#3}
           \mathop {\vtop {\m@th \ialign {##\crcr $\hfil \displaystyle {#3}\hfil $%
                              \crcr \noalign {\kern 3\p@ \nointerlineskip }\upbracketfill {#1}{#2}
                              \crcr \noalign {\kern 3\p@ }}}}\limits}
\def\upbracketfill#1#2{$\m@th \setbox \z@ \hbox {$\braceld$}
                  \edef\@bracketheight{\the\ht\z@}\bracketend{#1}{#2}
                  \leaders \vrule \@height #1 \@depth \z@ \hfill
                  \leaders \vrule \@height #1 \@depth \z@ \hfill \bracketend{#1}{#2}$}
\def\bracketend#1#2{\vrule height #2 width #1\relax}
\def\downbracketfill#1#2{$\m@th \setbox \z@ \hbox {$\braceld$}
                  \edef\@bracketheight{\the\ht\z@}\downbracketend{#1}{#2}
                  \leaders \vrule \@height #1 \@depth \z@ \hfill
                  \leaders \vrule \@height #1 \@depth \z@ \hfill
\downbracketend{#1}{#2}$}
\def\downbracketend#1#2{\vrule depth #2 width #1\relax}
\makeatother
\begin{document}
\pagenumbering{arabic}

\title{Spectral density of the non-central correlated Wishart ensembles}
\author{Vinayak}
\email{vinayaksps2003@gmail.com}
\affiliation{Instituto de Ciencias F\' isicas, Universidad Nacional Aut\' onoma de M\' exico, C.P. 62210 Cuernavaca, M\' exico}
\begin{abstract}

Wishart ensembles of random matrix theory have been useful in modeling positive definite matrices encountered in classical and quantum chaotic systems. We consider nonzero means for the entries of the constituting matrix $\mathbb{A}$ which defines the correlated Wishart matrix as $\mathbb{W}=\mathbb{AA}^{\dagger}$, and refer to the ensemble of such Wishart matrices as the non-central correlated Wishart ensemble (nc-CWE). We derive the Pastur self-consistent equation which describes the spectral density of nc-CWE at large matrix dimension. 
\end{abstract}
\pacs{02.50.Sk, 05.45.Tp, 89.90.+n}

\maketitle
\renewcommand*\thesection{\Roman{section}}
\renewcommand*\thesubsection{\thesection.\Roman{subsection}}

\section{Introduction}\label{Intro}
Random matrix theory (RMT) is no longer a specialized topic but its applications in a vast domain of science and advent of naive techniques made it important  in active area of research not only in physics \cite{Mehta, Brody81, TGW:98} and mathematics \cite{Wilks,Muirhead} but also in various other scientific disciplines \cite{Potters:2005, Seba:03, Sismology, diverseAT, bio}. The Wishart model for the correlation matrices \cite{Wishart}, introduced long way back in 1928, is probably the origin of RMT. In recent research, this model has gained much attention incorporating various generalizations in trend purposefully to model positive definite matrices encountered in classical or quantum chaotic systems. For instance the Wishart model, which incorporates actual correlations \cite{marchenko, Silverstien, SenM, MousSimon, Burda:2005, Baik:2005, vp2010,guhr1, guhr2}, gives a better platform to understand the underlying correlations in quantitative finance \cite{Finance1, Finance2, Finance3, Finance4} and also for practical statistical signal processing applications, including synthetic aperture radar, extra-solar planet detection, and multi-antenna wireless communications \cite{Muller:Review}. Ensemble of such Wishart matrices are known the correlated Wishart ensemble (CWE). Similarly, some other generalizations like using fixed-trace Wishart matrices while modeling the density matrices in quantum entanglement problems \cite{Fixed-Tr}, or power-map deformation of Wishart matrices \cite{vrt:2013, VBPS:2014} in the context of short time series analysis of multivariate systems, have also been useful. 

In a general sense the Wishart model may be defined as $\mathbb{W}=\mathbb{AA}^{\dagger}$ where $\mathbb{A}$ is of dimension $N\times T$. The matrix entries $A_{j\nu}$, for $1\le j\le N$ and $1\le \nu\le T$, are Gaussian variables with mean $\mu_{j\nu}$, variance $\sigma^{2}$ and with correlations, $\xi_{jk}$, between the $j$'th and $k$'th  rows of $\mathbb{A}$. In a usual set up where $\mu_{j\nu}=0$ and $\xi$ is diagonal with $1$, this model defines Wishart or Laguerre ensembles (WE) where a lot is known in terms of Laguerre polynomials for the eigenvalue statistics \cite{Laguerre, APSG}. If the off-diagonal terms of $\xi$ are not $0$, then the model defines the CWE. Using Dyson's classification of invariant ensembles \cite{Mehta, Potter:book}, the three invariant CWEs can be defined as the correlated Wishart orthogonal ensembles (CWOE), correlated Wishart unitary ensembles (CWUE) and correlated Wishart symplectic ensembles (CWSE). In this paper paper we consider rather a simple generalization for the all three invariant CWEs using $\mu_{j\nu}\ne 0$ which defines the non-central correlated Wishart ensembles (nc-CWE). Predating the Gaussian ensembles of RMT \cite{Mehta,Potter:book}, such non-central matrices were introduced in mathematical statistics to better the so-called {\it null hypothesis} supplied by CWE \cite{TWAnderson}. However, the non-central Wishart ensembles (nc-WE) have been revisited recently in the context of signal processing \cite{MIMO} and in mathematical statistics \cite{nc-WE1,nc-WE2,Forrester}. In physics, however, nc-CWEs have been used recently in the context of density matrices \cite{vmarko} remarking that the zero-mean condition is a priory not valid for the density matrices. It is also worth mentioning that the nonzero mean condition has also been studied for the Gaussian ensembles \cite{Gaussian, Forrester,ap81}.

For CWE, the spectral density is known in terms of a Pastur self-consistent equation \cite{marchenko,Silverstien, SenM, Burda:2005,vp2010} which is valid for large $N$ and $T$ with finite ratio $N/T=\kappa$. For $\xi=\mathbf{1}_{N}$, where $\mathbf{1}_{N}$ is $N\times N$ identity matrix, the Pastur equation yields the famous Mar\v cenko Pastur density for WE. For finite $N$ and $T$, CWE poses a serious difficulty and thus exact result is known only for the spectral density \cite{MousSimon, guhr1, guhr2} while the two-point spectral correlation is known only asymptotically \cite{vp2010} for large matrices. The Pastur equation, however, has never been investigated for nc-CWE and perhaps even for nc-WE. Our focus in this paper is to obtain the Pastur equation using the binary correlation method \cite{Brody81, ap81, vp2010, vin2013, vinLuis:2014} and investigate some important features like how do nonzero means and correlations affect the ensemble-averaged bulk density and the ensemble-averaged mean positions of the eigenvalues separated from the bulk. The latter has been important in RMT applications \cite{Bouchaud:2009}.

The paper is organized as follows. In the next section \ref{Prelim} we will describe the model and fix our notations. In Sec. \ref{LoopEq} we will derive the loop equation. In Sec. \ref{CWE} we re-derive the Pastur equation by solving the loop equation for the CWE case and discuss analytical results for the separated eigenvalues as derived in Ref. \cite{vp2010}. In the first part of Sec. \ref{nc-WE} we will specialize in deriving the Pastur equation for nc-WE. In the second part of the same section we will derive the ensemble-averaged mean position of the separated eigenvalues and discuss their universality with the CWE case. Finally in the third part, with Monte-Carlo simulations we will illustrate our analytical result for the bulk density. Similarly, in Sec. \ref{nc-CWE} we will generalize the method of Secs. \ref{CWE} and \ref{nc-WE} and derive the Pastur equation for nc-CWE in the first part. Next, we will discuss about the separated eigenvalues and the bulk density respectively in the second and third part. Finally, we summarize our work with discussions in Sec. \ref{conclusion}.

\section{Preliminaries}\label{Prelim}
The model we are interested in is defined as
\begin{equation}
\mathbb{W}=\mathbb{AA}^{\dagger}/T,
\end{equation}
where $\mathbb{A}$ is $N\times T$ and
\begin{equation}
\mathbb{A}=\xi^{1/2}\mathbf{A}+\mathsf{B},
\end{equation}
so that, $\overline{\mathbb{A}}=\mathsf{B}$ and $\overline{\mathbb{W}}=v_{1}^{2}\xi+\mathsf{BB}^{\dagger}/T$ where $\xi$ is the $N\times N$ positive definite fixed (nonrandom) matrix defining correlations between rows of $\mathbb{A}$ and $\mathsf{B}$ is the $N\times T$ fixed matrix which representing the ensemble averaged $\mathbb{A}$. Here we have used an {\it overbar} for the ensemble averaging. Indeed, $\mathbf{A}$ is the random matrix where the matrix entries $A_{jk}$ are real Gaussian variables with mean $0$ and variance $v_{1}^{2}$ for nc-CWOE where the Dyson index $\beta=1$. Similarly for the nc-CWUE, $\beta=2$ and we consider $A_{jk}=A^{(1)}_{jk}+\ii A^{(2)}_{jk}$ where $\mathsf{A}^{(1)}$ and $\mathsf{A}^{(2)}$ statistically equivalent but independent Gaussian matrices with mean $0$ and variance $v_{2}^{2}$. Finally, for the nc-CWSE symplectic ensembles $\beta=4$ and $\mathsf{A}$ is composed of $4$ statistically equivalent but independent Gaussian matrices with mean $0$ variance $v_{4}^{2}$ written in terms of $\mathbf{1}_{2}$ and two-dimensional matrix representative of quaternion units $\tau_{\gamma}$ where $\gamma=1,...,3$. Then $\mathbb{A}^{\dagger}$ is the transpose, Hermitian conjugate and dual of $\mathbb{A}$ respectively for $\beta=1,2,$ and $4$.The joint probability density of the matrix elements of $\mathbf{A}$ is given by the Gaussian probability measure,
\begin{equation}\label{MGaussian}
\mathcal{P}(\mathbf{A})\propto \exp\left[
-\text{Tr}\frac{\mathbf{A}\mathbf{A}^{\dagger}}{2v_{\beta}^{2}}
\right].
\end{equation}
Since variance supplies the scale for the statistics, we fix the scale as $v_{\beta}^{2}=\sigma^{2}\beta^{-1}$ \cite{MonFrench}. With out loss of generality we consider $T\ge N$.  
% for which $\mathbb{W}$ will always be a full rank matrix. 

We use the binary correlation method in order to obtain the ensemble-averaged spectral density, $\overline{\rho}_{\mathbb{W}}(\lambda)$. In this method it is convenient to deal with the Stieltjes transform or the resolvent of the density while the resolvent, $\overline{\mathbf{g}}_{\mathbb{W}}(z)$, is defined as
\begin{equation}
\overline{\mathbf{g}}_{\mathbb{W}}(z)=\overline{\langle \,\left(z\mathbf{1}_{N}-\mathbb{W}\right)^{-1}\,\rangle_{N}},
\end{equation}
where $z=\lambda\pm \ii \epsilon$ for positive infinitesimal $\epsilon$ and the angular brackets stand for the spectral averaging, e.g. $\langle \mathbf{H} \rangle_{K}= K^{-1}\, \text{tr}\, \mathbf{H}$ for $K\times K$ dimensional $\mathbf{H}$. Then $\overline{\rho}_{\mathbb{W}}(\lambda)$ can be determined via the relation
\begin{equation}\label{rhoG}
\overline{\rho}_{\mathbb{W}}(\lambda)=\lim_{\epsilon\to 0} \frac{\mp}{\pi} \Im \, \overline{\mathbf{g}}_{\mathbb{W}}(z).
\end{equation}
In order to calculate $\overline{\mathbf{g}}_{\mathbb{W}}(z)$ we use the moment expansion:
\begin{equation}\label{mom1}
\overline{\mathbf{g}}_{\mathbb{W}}(z)=\sum_{n=0}^{\infty}\frac{{\bf \overline{m}}_{n}}{z^{n+1}},
\end{equation}
where $\overline{{\bf m}}_{n}$ is the $n$'th moment of $\overline{\rho}_{\mathbb{W}}(\lambda)$ defined as
\begin{equation}\label{mom2}
\overline{{\bf m}}_{n}=\int d\lambda\, \lambda^{n}\overline{\rho}_{\mathbb{W}}(\lambda)=\overline{\langle \mathbb{W} ^{n}\rangle}_{N}. 
\end{equation}

In principle, the problem is solved once we obtain a closed form of $\overline{\mathbf{g}}_{\mathbb{W}}(z)$. As in \cite{vp2010}, we could have started the moment expansion (\ref{mom1}, \ref{mom2}) to obtain $\overline{\mathbf{g}}_{\mathbb{W}}(z)$, but due to the additional term $\mathsf{B}$ the expansion results nontrivial combinations of $\mathsf{A}$ and $\mathsf{B}$. Further complications will arise in the ensemble averaging of this series with respect to the jpd (\ref{MGaussian}). We simplify the problem regarding the ensemble averaging first by using the trick of linearization \cite{Nowak-Nowak}. Following Ref. \cite{Nowak-Nowak}, we define 
\begin{equation}
\mathsf{X}=\frac{1}{\sqrt{T}}\left[
\left(\begin{matrix}
\mathbf{0} & \mathsf{A}\\
\mathsf{A}^{\dagger} & \mathbf{0}
\end{matrix}\right)
+\left(
\begin{matrix}
\mathbf{0} & \mathsf{B}\\
\mathsf{B}^{\dagger} & \mathbf{0}
\end{matrix}
\right)
\right],
\end{equation}
where we replaced $\xi^{1/2}\mathbf{A}$ by $\mathsf{A}$. In the following, we use $(N+T)\times(N+T)$ matrices defined as
\begin{equation}
\tilde{\mathsf{A}}= \left(\begin{matrix}
\mathbf{0} & \mathsf{A}\\
\mathsf{A}^{\dagger} & \mathbf{0}
\end{matrix}\right), ~~~ \tilde{\mathsf{B}}=\left(
\begin{matrix}
\mathbf{0} & \mathsf{B}\\
\mathsf{B}^{\dagger} & \mathbf{0}
\end{matrix}
\right).
\end{equation}
Note that the eigenvalues of $\mathsf{X}^{2}$ coincides with those of $\mathbb{W}$ with a two-fold degeneracy for each. We define resolvent, $\overline{\mathbf{g}}_{\mathsf{X}}(u)$, for the spectral density $\overline{\rho}_{\mathsf{X}}(y)$ of $\mathsf{X}$, as
\begin{equation}\label{gx}
\overline{\mathbf{g}}_{\mathsf{X}}(u)=\overline{\langle (\mathsf{U}-\mathsf{X})^{-1} \rangle}_{N+T}, ~\text {where}~
\mathsf{U}=u\mathbf{1}_{N+T},
\end{equation}
and $u=y\pm\ii\epsilon$. In what follows, we calculate $\overline{\mathbf{g}}_{\mathbb{W}}(z)$ using $\overline{\mathbf{g}}_{\mathsf{X}}(u)$. For the first and so on, moments of $\overline{\rho}_{\mathsf{X}}$ are related with the moments of $\overline{\rho}_{\mathbb {W}}$ via 
\begin{equation}\label{gwgu}
z \overline{\mathbf{g}}_{\mathbb{W}}(z)-1=\frac{N+T}{2N}\left(u(z)\overline{\mathbf{g}}_{\mathsf{X}}(u(z))-1\right),
\end{equation}
and $u^{2}=z$. Below, we will calculate the ensemble average of 
\begin{equation}\label{GLX}
\mathsf{G}^{(\mathsf{X})}_{\mathsf{L}}(u)=\mathsf{L} \left(\mathsf{U-X}\right)^{-1},
\end{equation}
where 
\begin{equation}
\mathsf{G}^{(\mathsf{X})}=\left(
\begin{matrix}
G_{11} & G_{12}\\
G_{21} & G_{22}
\end{matrix}
\right).
\end{equation}
$G_{jj}$ are square blocks, of dimensions $N\times N$ and $T\times T$ respectively for $j=1$ and $2$, and $G_{12}$ and $G_{21}$ are rectangular blocks of dimensions $N\times T$ and $T\times N$ respectively and $\mathsf{L}$ is an $(N+T)\times (N+T)$ arbitrary fixed matrix. $\mathsf{L}=\mathbf{1}_{N+T}$ gives $\mathsf{G}^{(\mathsf{X})}$ which on the spectral averaging yields $\overline{\langle\mathsf{G}^{(\mathsf{X})}}(u)\rangle_{N+T}=\overline{\mathbf{g}}_{\mathsf{X}}(u)$. Finally, we define the ratio
\begin{equation}
\kappa=N/T.
\end{equation}

\section{The loop equation}\label{LoopEq}
We notice that the large-$u$ expansion of $\mathsf{G}^{(\mathsf{X})}_{\mathsf{L}}(u)$ has non-trivial combinations of $\tilde{\mathsf{A}}$ and $\tilde{\mathsf{B}}$. Since $\tilde{\mathsf{B}}$ is a fixed matrix, we may use
\begin{equation}\label{KUB}
\mathsf{K}= \left(\mathsf{U}-\tilde{\mathsf{B}}\right)^{-1},
\end{equation}
and expand $\mathsf{G}^{(\mathsf{X})}_{\mathsf{L}}(u)$ the for small $\mathsf{K}$ (or equivalently for large $u$). It is worth mentioning that this trick has been used in the context of non-central Gaussian ensembles in Ref. \cite{ap81}. Then the large-$u$ expansion of Eq. (\ref{GLX}) can be written as
\begin{equation}
\mathsf{G}^{(\mathsf{X})}_{\mathsf{L}}(u)= \mathsf{LK}\sum_{n=0}^{\infty} \left(\tilde{\mathsf{A}}\mathsf{K}\right)^{n}.
\end{equation}
Since $A_{jk}$ are centered at $0$, the odd-$n$ terms of the above expansion are identically $0$ on the ensemble averaging. Thus the ensemble-averaged series reduces to
\begin{equation}\label{series}
\overline{\mathsf{G}^{(\mathsf{X})}_{\mathsf{L}}}(u)= \mathsf{LK} +\mathsf{LK}\overline{\tilde{\mathsf{A}}\mathsf{G}^{(\mathsf{X})}\tilde{\mathsf{A}}\mathsf{G}^{(\mathsf{X})}}.
\end{equation}
In order to perform the ensemble averaging for the remaining terms we use the jpd (\ref{MGaussian}) with $\mathsf{A}=\xi^{1/2}\mathbf{A}$ and derive the following exact identities, valid for arbitrary fixed $\Phi$ and $\Psi$,
\begin{eqnarray}
\label{Id-1}
\frac{1}{T} \overline{\mathsf{A}\Phi\mathsf{A}^{\dagger}\Psi}&=& \sigma^{2} \langle\Phi\rangle_{T} \xi \Psi,\\
\label{Id-2}
\frac{1}{T} \overline{\mathsf{A}^{\dagger}\Phi\mathsf{A}\Psi}&=& \sigma^{2} \langle\xi\Phi\rangle_{T} \Psi,\\
\label{Id-3}
\overline{\mathsf{A}\Phi\mathsf{A}\Psi}&=&\frac{(2-\beta)\sigma^{2}}{\beta}\Psi\tilde{\Phi},
\end{eqnarray}
where $\tilde{\Phi}= \Phi^{t}$, for $\beta=1$ where $\Phi^{t}$ is the transpose of $\Phi$, $\tilde{\Phi}= \Phi$ for $\beta=2$ and $\tilde{\Phi}= -\tau_{2}\Phi^{t}\tau_{2}$. As the identities suggest, we consider only the terms resulting form the binary associations of $\mathsf{A}$ with $\mathsf{A}^{\dagger}$ and avoid terms resulting from the binary associations of $\mathsf{A}$ with $\mathsf{A}$. With the help of these identities we calculate only the leading order terms of the series in Eq. (\ref{series}). We find
\begin{equation}\label{avSeries}
\overline{\mathsf{G}^{(\mathsf{X})}_{\mathsf{L}}}=\mathsf{LK}+\mathsf{LK}\Sigma \overline{\mathsf{G}^{(\mathsf{X})}},
\end{equation}
where the equality is valid only in the leading order and 
\begin{equation}\label{Sigma}
\Sigma=\sigma^{2}\left(
\begin{matrix}
\xi\overline{\langle G_{22}\rangle}_{T} & \mathbf{0}\\
 \mathbf{0} & \kappa\overline{\langle\xi G_{11}\rangle}_{N}\mathbf{1}_{T}
\end{matrix}
\right).
\end{equation}
In the derivation of the above Eqs. (\ref{avSeries}, \ref{Sigma}) we have avoided binary associations across the traces as those also result terms of $O(N^{-1})$. Substituting now $\mathsf{L}\to \mathsf{L}(\mathbf{1}_{N}-\mathsf{K}\Sigma)^{-1}$ in Eq. (\ref{avSeries}), and then Eq. (\ref{KUB}), we finally get
\begin{equation}\label{GLSigma}
\overline{\mathsf{G}^{(\mathsf{X})}_{\mathsf{L}}}(u)=\mathsf{L}\left(\mathsf{U}-\tilde{\mathsf{B}}-\Sigma\right)^{-1}.
\end{equation}

In order to calculate the inverse of the matrix in the right-hand-side (r.h.s) of the Eq. (\ref{GLSigma}), we use the Schur decomposition. For instance, using $\mathsf{M}=\mathsf{U}-\tilde{\mathsf{B}}-\Sigma$, we may write
\begin{equation}\label{mat-inv}
\mathsf{M}^{-1}=\left(
\begin{matrix}
\mathbf{a} & \mathbf{b}\\
\mathbf{c} & \mathbf{d}
\end{matrix}
\right)^{-1}=
\left(
\begin{matrix}
\mathsf{S}^{-1} & -\mathsf{S}\mathbf{b}\mathbf{d}^{-1}\\
-\mathbf{d}^{-1}\mathbf{c}\mathsf{S}^{-1} & (\mathbf{d}-\mathbf{c}\mathbf{a}^{-1}\mathbf{b})^{-1}
\end{matrix}
\right),
\end{equation}
where $\mathsf{S}=\mathbf{a}-\mathbf{b}\mathbf{d}^{-1}\mathbf{c}$ and 
\begin{eqnarray}
&&\mathbf{a}= u\mathbf{1}_{N}-\sigma^{2}\xi \overline{g}_{22}, ~~\mathbf{b}=-\frac{1}{\sqrt{T}}\mathsf{B},\nonumber\\
&&\mathbf{c}=-\frac{1}{\sqrt{T}}\mathsf{B}^{\dagger}, ~~\mathbf{d}=(u-\sigma^{2}\kappa\,\overline{g}_{11;\xi})\mathbf{1}_{T}.
\end{eqnarray}
We have used here more general spectral averaged quantities, defined as
\begin{equation}\label{Spec-Ave}
\overline{g}_{jj;\mathcal{L}}=\langle\mathcal{L} \overline{G}_{jj}\rangle_{K},
\end{equation}
with $\mathcal{L}$ as an arbitrary fixed matrix of the dimension $N\times N$ and $T\times T$ and $K$ is $N$ and $T$, respectively for $j=1$ and $2$. For example, the spectral-averaged quantity $\overline{g}_{22}$ is obtained by using $\mathcal{L}=\mathbf{1}_{T}$ in definition (\ref{Spec-Ave}) for the corresponding  upper diagonal-block matrix of the r.h.s. of Eq. (\ref{Sigma}). Similarly, for the lower diagonal-block matrix we have used $\mathcal{L}=\xi$: $\overline{g}_{11;\xi}=\langle\xi \overline{G}_{11}\rangle_{N}$.

Next, we use $\mathsf{L}=\mathbf{1}_{N+T}$ in Eq. (\ref{GLSigma}) and compute $\overline{\mathbf{g}}_{\mathsf{X}}(u)$ using Eq. (\ref{gx}). We get
\begin{equation}\label{gu-g11g22}
\overline{\mathbf{g}}_{\mathsf{X}}(u)=\langle(\overline{g}_{11}\mathbf{1}_{N}\oplus \overline{g}_{22}\mathbf{1}_{T})\rangle_{N+T},
\end{equation} 
where $\oplus$ stands for the direct sum and
\begin{eqnarray}
\label{g11}
\overline{g}_{11}&=&
\llan
%\left[
\frac{1}
{u\mathbf{1}_{N}-\sigma^{2}\xi\,\overline{g}_{22}-\frac{\zeta}{(u-\sigma^{2}\kappa\overline{g}_{11;\xi})}}
%\right]^{-1}
\rran_{N},\\
\label{g22}
\overline{g}_{22}&=&
\llan
%\left[
\frac{1}
{(u-\sigma^{2}\kappa\overline{g}_{11;\xi})\mathbf{1}_{T}-\frac{1}{T}\mathsf{B}^{\dagger}\,(u-\sigma^{2}\xi\,\overline{g}_{22})^{-1}\mathsf{B}}
%\right]^{-1}
\rran_{T}.\nonumber\\
\end{eqnarray}
In the above equation we have introduced a positive definite matrix $\zeta=\mathsf{BB}^{\dagger}/T$. Note that $\overline{\mathbf{g}}_{\mathbb{W}}(z)$ can be obtained by calculating $\overline{\mathbf{g}}_{\mathsf{X}}(u)$ using the relation (\ref{gu-g11g22}) and then using the relation (\ref{gwgu}).

\section{Pastur Equation for CWE}\label{CWE}

For our model, $B_{jk}=0$ defines the CWE. The spectral density of CWE has been derived by several authors \cite{marchenko, Silverstien, SenM, Burda:2005, vp2010} using different techniques. As mentioned before, for large $N$ and $T$ with finite ratio $\kappa$, the spectral density is known in terms of the Pastur self-consistent equation. Below we give an alternative method to obtain the Pastur density for CWE by solving the loop equation (\ref{g11}, \ref{g22}). 

We first note that in this case Eqs. (\ref{g11}, \ref{g22}) reduce to 
\begin{equation}\label{g11g22-CWE}
\overline{g}_{11}(u)=\llan(u\mathbf{1}_{N}-\sigma^{2}\xi\, \overline{g}_{22})^{-1}\rran,~~\overline{g}_{22}(u)=(u-\sigma^{2}\kappa \overline{g}_{11;\xi})^{-1},
\end{equation}
where 
\begin{equation}\label{g11xi-CWE}
\overline{g}_{11;\xi}(u)=\llan\xi\,(u\mathbf{1}_{N}-\sigma^{2}\xi\, \overline{g}_{22})^{-1}\rran
=\frac{u\overline{g}_{11}(u)-1}{\sigma^{2}\overline{g}_{22}(u)}.
\end{equation}
We may also write $\overline{g}_{22}$ as
\begin{equation}
u\overline{g}_{22}(u)=1+\sigma^{2}\kappa \overline{g}_{11;\xi}(u) \overline{g}_{22}(u).
\end{equation}
Using the second equality of Eq. (\ref{g11xi-CWE}) in the above equation we get
\begin{eqnarray}\label{g11g22}
\overline{g}_{22}
&=&
\frac{\kappa \,u\overline{g}_{11}+1-\kappa}{u}.
\end{eqnarray}
This is a very useful equation because not only it establishes a linear relation between $\overline{g}_{11}$ and $\overline{g}_{22}$ that we need to solve the loop equation but also when inserted in Eqs. (\ref{gwgu},\ref{gu-g11g22}) it leads to another useful identity:
\begin{equation}\label{gzgu}
z\overline{\mathbf{g}}_{\mathbb{W}}(z)=u(z) \overline{g}_{11}(u(z)). % invariant of \sigma^{2}
\end{equation}
As it will shown ahead, the above two relations (\ref{g11g22}, \ref{gzgu}) are also valid for nc-WE and nc-CWE. Finally, we use these two relations in Eq. (\ref{g11g22-CWE}) with $z=u^{2}$ and obtain the Pastur density for CWE:
\begin{equation}\label{Pastur-CWE}
\overline{\mathbf{g}}_{\mathbb{W}}(z)=\llan\left[z\mathbf{1}_{N}-\sigma^{2}(1-\kappa+z\kappa\overline{\mathbf{g}}_{\mathbb{W}}(z))\xi\right]^{-1}\rran_{N}.
\end{equation}
As noted in Ref. \cite{vp2010} that this result is independent of the Dyson-index $\beta$ because of the scaling $v_{\beta}^{2}=\sigma^{2}/\beta$. The same holds true for the other Pastur equations we derive below. 

For a nontrivial spectrum of $\xi$ the analytic solution is complicated. Thus, this equation has to be solved numerically. To this end an efficient algorithm is discussed in Ref. \cite{vp2010} where various cases of $\xi$ have been worked out. However, analytically we can solve the Pastur equation when it is quadratic. For instance, consider the $\xi_{jk}=\delta_{jk}$. For this choice the Pastur equation (\ref{Pastur-CWE}) yields the resolvent 
\begin{equation}\label{GMP}
\overline{\mathbf{g}}_{\mathbb{W}}(z)=\frac{z-\sigma^{2}(1-\kappa)-\sqrt{(z-\sigma^{2}(1-\kappa))^{2}-4z\kappa\sigma^{2}}}{2\kappa z\sigma^{2}},
\end{equation}
where we have considered the negative sign so that $\overline{\mathbf{g}}_{\mathbb{W}}(z)$ behaves as $z^{-1}$ for large $z$. Next, the inverse transform (\ref{rhoG}) of this resolvent gives the famous Mar\v cenko Pastur density:
\begin{equation}\label{MPlaw}
\overline{\rho}_{\text{MP}}(\lambda)=\frac{\sqrt{(\lambda_{+}-\lambda)(\lambda-\lambda_{-})}}{2\pi \kappa\sigma^{2} \lambda}%+\delta(\lambda-\overline{\lambda}),
\end{equation}
where $\lambda_{\pm}=\sigma^{2}(\sqrt{\kappa}\pm1)^{2}$. 

It has been shown in Ref. \cite{vp2010} that Eq. (\ref{Pastur-CWE}) can also be solved for the equal-cross-correlation matrix model, viz. $\xi_{jk}=\delta_{jk}+(1-\delta_{jk})\, \mu_{0}^{2}$. Notice that in this case $\xi$ is diagonal plus a rank-$1$ matrix. Thus for $N\mu_{0}^{2}>\sqrt{\kappa}$ the spectral density we find is composed of a bulk and a separated eigenvalue provided $N\mu_{0}^{2}>\sqrt{\kappa}$:
\begin{equation}\label{den-eqcor}
\overline{\rho}_{\mathbb{W}}(\lambda)=\overline{\rho}_{0}(\lambda)+N^{-1}\delta(\lambda-\overline{\lambda}_{N}).
\end{equation}
The bulk density, $\overline{\rho}_{0}(\lambda)$, described by the Mar\v cenko Pastur law (\ref{MPlaw}) with a rescaled variance $\sigma^{2}(1-\mu_{0}^{2})$. The ensemble-averaged position of the separated eigenvalues, $\overline{\lambda}$, is given by
\begin{eqnarray}\label{Ebar-CWE}
\overline{\lambda}_{N}&=&\sigma^{2}\frac{[(N-1)\mu_{0}^{2}+1)][(N-\kappa)\mu_{0}^{2}+\kappa]}{N\mu_{0}^{2}}\nonumber\\
&\simeq& \sigma^{2}\frac{(N\mu_{0}^{2}+1)(N\mu_{0}^{2}+\kappa)}{N\mu_{0}^{2}}.
\end{eqnarray}
A simple generalization of the equal-cross-correlation matrix is a block diagonal matrix, which is again diagonal plus a finite-rank matrix. In this case the above result can be easily generalized for other separated eigenvalues. However, it has been shown in Ref. \cite{vp2010} that even for more complicated $\xi$ analytic result for the $k$'th separated eigenvalue $\overline{\lambda}_{k}$ can be written as
\begin{equation}\label{Ebark-CWE}
\overline{\lambda}_{k}=\sigma^{2}\lambda^{(\xi)}_{k}\left(
1-\kappa+\lambda^{(\xi)}_{k}\llan\,\mathbb{Q}_{k}(\lambda^{(\xi)}_{k}\,\mathbf{1}_{N}-\xi)^{-1}\rran_{N}
\right),
\end{equation}
where $\lambda^{(\xi)}_{k}$ is the $k$'th eigenvalue of $\xi$ and $\mathbb{Q}_{k}=\mathbf{1}_{N}-|k\rangle\,\langle k|$ is the projection operator to the $k$'th eigenstate $|k\rangle$ of $\xi$.

\section{Pastur equation for nc-WE}\label{nc-WE}
nc-WE is perhaps the simplest case next to WE or CWE. The nc-WEs have already been addressed in Ref. \cite{Forrester} using different methods. Since the Pastur equation has never been given explicitly, below we derive the Pastur equation for nc-WEs.  

We begin with using $\xi=\mathbf{1}_{N}$ in  Eqs. (\ref{g11}, \ref{g22}) which results
\begin{eqnarray}\label{g11-NCW}
\overline{g}_{11}&=&\llan 
\frac{1}
{(u-\sigma^{2}\overline{g}_{22})\mathbf{1}_{N}-\frac{\zeta}{u-\sigma^{2}\kappa\overline{g}_{11}}}
\rran_{N},
\\
\label{g22-NCW}
\overline{g}_{22}&=&
\llan
\frac{1}
{(u-\sigma^{2}\kappa\overline{g}_{11})\mathbf{1}_{T}-\frac{\eta}{u-\sigma^{2}\overline{g}_{22}}}
\rran_{T},
\end{eqnarray} 
where in the second equality we have used $\mathsf{B}^{\dagger}\mathsf{B}/T=\eta$. Notice that except for the zeros, $\zeta$ and $\eta$ both have the spectrum. As mentioned above, Eqs. (\ref{g11g22}, \ref{gzgu}) also hold here. To show this we first write
\begin{eqnarray}\label{g11g22-ncWE}
\overline{g}_{22}
&=&
\frac{1}{T}\sum_{j=1}^{N}
\left[
\frac{u-\sigma^{2}\overline{g}_{22}}
{(u\mathbf{1}_{N}-\sigma^{2}\kappa\overline{g}_{11})(u-\sigma^{2}\overline{g}_{22})-\lambda^{(\zeta)}_{j}}
\right]
\nonumber\\
&+&
\frac{(1-\kappa)}{(u\mathbf{1}_{N}-\sigma^{2}\kappa\overline{g}_{11})}.
\end{eqnarray}
Next, we use Eq. (\ref{g11-NCW}) in the above equality and obtain (\ref{g11g22}) which consequently implies the relation (\ref{gzgu}). Finally, we use the relations (\ref{g11g22},\ref{gzgu}) and substitute $u^{2}=z$ to simplify the loop equation (\ref{g11-NCW}) into the self-consistent equation for $\overline{\mathbf{g}}_{\mathbb{W}}(z)$. This method yields the Pastur equation for nc-WE:
\begin{equation}\label{Pastur-NcW}
\overline{\mathbf{g}}_{\mathbb{W}}(z)=
\llan
\frac{1}{
[z-\sigma^{2}(1-\kappa+z\kappa\overline{\mathbf{g}}_{\mathbb{W}}(z))]\mathbf{1}_{N}-\frac{\zeta}{1-\sigma^{2}\kappa\,\overline{\mathbf{g}}_{\mathbb{W}}(z)}
}
\rran_{N}.
\end{equation}

If we set now $\zeta=0$, then we indeed get resolvent of the Mar\v cenko-Pastur density (\ref{GMP}). Otherwise, if we set $\sigma^{2}=0$ then it will give the resolvent corresponding the spectrum of $\zeta$. Like the Pastur equation for CWE, here as well, Eq. (\ref{Pastur-NcW}) depends on the spectrum of $\zeta$ and thus has be solved numerically when it has a non-trivial spectrum. Below we consider a rank-$1$ matrix $\mathsf{B}$ which closely related with the equal-cross correlation matrix model of the CWE. However, unlike CWE in this case the bulk density is not rescaled with variance but remains the same as for the WE (\ref{MPlaw}). Using the techniques of Refs. \cite{vp2010,ap81} we start with this simple choice to calculate the ensemble averaged position of the separated eigenvalues and generalize this result for the bulk density different from the Mar\v cenko Pastur density.

\subsection{Separation of Eigenvalues}
We begin with a simple choice for $\mathsf{B}$, viz.
\begin{equation}
B_{jk}=\mu.
\end{equation}
Then the only nonzero eigenvalue of $\zeta$, $\lambda^{(\zeta)}_{N}=N\mu^{2}$. In this case, from Eq. (\ref{Pastur-NcW}) we get 
\begin{eqnarray}\label{gncw-FinRank}
&&\overline{\mathbf{g}}_{\mathbb{W}}(z)=\overline{\mathbf{g}^{(0)}}(z)\nonumber\\
&+&
\frac{N^{-1}}{
z-\sigma^{2}(1-\kappa+z\kappa\overline{\mathbf{g}}_{\mathbb{W}}(z))-\frac{\lambda^{(\zeta)}_{N}}{1-\sigma^{2}\kappa\,\overline{\mathbf{g}}_{\mathbb{W}}(z)}}.
\end{eqnarray}
Here we have used
\begin{eqnarray}\label{g0-ncw}
\overline{\mathbf{g}^{(0)}}(z)&=&
\Big\langle \mathbb{Q}_{N}
%\frac{1}{
\Big[
z\mathbf{1}_{N}-\sigma^{2}(1-\kappa+z\kappa\overline{\mathbf{g}}_{\mathbb{W}}(z))\nonumber\\
&-&\frac{\zeta}{1-\sigma^{2}\kappa\,\overline{\mathbf{g}}_{\mathbb{W}}(z)}
\Big]^{-1}
%}
\Big\rangle_{N},
\end{eqnarray}
where $\mathbb{Q}^{(\zeta)}_{k}=\mathbf{1}_{N}-|k\rangle\langle k|$ and $|k\rangle\langle k|$ is the projection operator for the eigenstate $|k\rangle$ corresponding to the eigenvalue $\lambda^{(\zeta)}_{k}$. Solving Eq. (\ref{gncw-FinRank}), while ignoring the second term, we retrieve the Mar\v cenko-Pastur result (\ref{GMP}) for the bulk density while it is understood that the bulk density is normalized to $1-1/N$. However, in the the above equation we do not drop the term containing $\zeta$ and treat this term as for a general $\zeta$. The ensemble-averaged mean position of the separated eigenvalues can be identified from the pole in the second term of Eq. (\ref{gncw-FinRank}) as
\begin{equation}\label{pole-ncw}
\overline{\lambda}_{N}=\sigma^{2}(1-\kappa+\overline{\lambda}_{N}\kappa\overline{\mathbf{g}^{(0)}}(\overline{\lambda}_{N}))+\frac{\lambda^{(\zeta)}_{N}}{1-\sigma^{2}\kappa\,\overline{\mathbf{g}^{(0)}}(\overline{\lambda}_{N})},
\end{equation}
where we have used $\overline{\mathbf{g}^{(0)}}$ instead of $\overline{\mathbf{g}}$ and ignored $O(N^{-1})$ terms. Using this in Eq. (\ref{g0-ncw}) we obtain 
\begin{equation}\label{gEbar-ncw}
\overline{\mathbf{g}^{(0)}}(\overline{\lambda}_{N})=\frac{\Phi_{N}}{1+\sigma^{2}\kappa\Phi_{N}},
\end{equation}
where 
\begin{equation}
\Phi_{N}=\llan \mathbb{Q}_{N}(\lambda^{(\zeta)}_{N}\mathbf{1}_{N}-\zeta)^{-1}\rran_{N}.
\end{equation}
Next, by using Eq. (\ref{gEbar-ncw}) in Eq. (\ref{pole-ncw}), we obtain $\overline{\lambda}_{N}$. Following Ref. \cite{vp2010}, we can also generalize this result for the $k$'th separated eigenvalue, $\overline{\lambda}_{k}$ as
\begin{equation}
\overline{\lambda}_{k}=(1+\sigma^{2}\kappa\Phi_{k})[\sigma^{2}(1-\kappa)+\lambda^{(\zeta)}_{k}(1+\sigma^{2}\kappa\Phi_{k})]. %verified numerically
\end{equation}

The above result is of course different from that for the CWE (\ref{Ebark-CWE}). However, for the rank-one $\mathsf{B}$ this result gives
\begin{equation}\label{Sprt-NCW}
\overline{\lambda}_{N}=\frac{(N\mu^{2}+\sigma^{2})(N\mu^{2}+\sigma^{2}\kappa)}{N\mu^{2}},
\end{equation}
which is valid only if $N\mu^{2}>\sqrt{\kappa}\sigma^{2}$ otherwise the separated eigenvalue will be absorbed in the Mar\v cenko-Pastur-bulk. Interestingly, it also coincides with (\ref{Ebar-CWE}) for $\mu=\mu_{0}$ and $\sigma^{2}=1$. In Ref. \cite{vmarko}, this correspondence has been exploited without any analytical treatment for the nc-WE. There are the parameters chosen as $\mu=\sqrt{r/N}$ and $\sigma=(1-r)/\sqrt{N}$ in the n-WUE case and $\mu_{0}=\sqrt{r}$ and $\sigma=1/\sqrt{N}$ in the CWUE case. Indeed, for these parameters the two results (\ref{Ebar-CWE}) and (\ref{Sprt-NCW}) coincide in the leading order.

\subsection{The Bulk Density}

\begin{figure}
        \centering
               \includegraphics [width=0.5\textwidth]{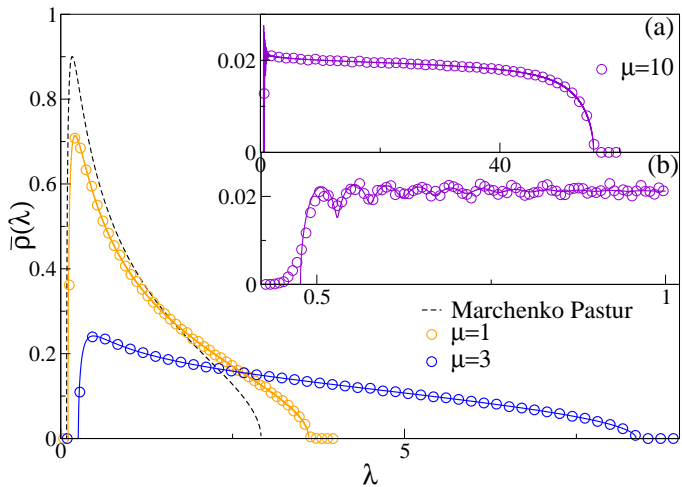}
                \caption{(Color online) Spectral density of the nc-WOE where $B_{j\nu}=\delta_{j\nu}\mu\sqrt{j}$ for $N=1024$, $T=2N$, $\sigma^{2}=1$, and $\mu=0,1,3$ and $10$. Solid lines in this figure represent the theory obtained from the numerical solution of \ref{Pastur-NcW} and open circles represent the histogram data obtained from the Monte-Carlo simulation of $\mathsf{C}$. The dashed line represent the Mar\v cenko-Pastur formula. In the inset (a) we show the density for $\mu=10$ while in (b) we show the density for $\lambda\le1$ with the same $\mu$.}
\label{Rank-N}
\end{figure}

It is important to point out that the Eq. (\ref{Pastur-NcW}) describes only the bulk density and not the density of the separated eigenvalues. Analytically it has been proved for CWUE that the density of the separated eigenvalues is described by a Gaussian distribution \cite{Baik:2005} and numerically the same is found to be valid for the nc-WUE case \cite{vmarko}. To obtain the bulk density for a non-trivial $\zeta$, Eq. (\ref{Pastur-NcW}) has to be solved numerically. We thus use the Newton's method described in Ref. \cite{vp2010} to solve Eq. (\ref{Pastur-NcW}). We consider
\begin{equation}
f(\overline{\mathbf{g}^{(n)}_{\mathbb{W}}}(z))-\overline{\mathbf{g}^{(n)}_{\mathbb{W}}}(z)=0, 
\end{equation}
at given $z$ where $f(\overline{\mathbf{g}^{(n)}_{\mathbb{W}}}(z))$ is the r.h.s. of Eq. (\ref{Pastur-NcW}) for $\overline{\mathbf{g}^{(n)}_{\mathbb{W}}}(z)$, and $n$ represent the iteration-number starting from $0$ with an initial guess $\overline{\mathbf{g}^{(0)}_{\mathbb{W}}}(z)$. %Next, we iterate the sequnce changing $\overline{\mathbf{g}^{(n)}_{\mathbb{W}}}(z)$ at the $n$'th iteration by
%\begin{eqnarray}
%\delta \overline{\mathbf{g}^{(n)}_{\mathbb{W}}}(z) &=&
%\frac{\mathbf{g}^{(n)}_{\mathbb{W}}(z)-f(\overline{\mathbf{g}^{(n)}_{\mathbb{W}}}(z))}{f'(\overline{\mathbf{g}^{(n)}_{\mathbb{W}}}(z))-1}\nonumber\\
%f'(\overline{\mathbf{g}^{(n)}_{\mathbb{W}}}(z))&=&
%\sigma^{2}\kappa
%\Big\langle
%(z\mathbf{1}_{N}+\zeta\,(1-\sigma^{2}\kappa\overline{\mathbf{g}^{(n)}_{\mathbb{W}}})\nonumber\\
%&\times&\frac{1}{[z-\sigma^{2}(1-\kappa+z\kappa\overline{\mathbf{g}}_{\mathbb{W}}(z))-\frac{\zeta}{1-\sigma^{2}\kappa\,\overline{\mathbf{g}}_{\mathbb{W}}(z)}]^{2}}
%\Big\rangle
%\end{eqnarray}

To illustrate the result (\ref{Pastur-NcW}) we use $B_{j\nu}=\delta_{j\nu}\mu\sqrt{j}$, for $0\le j\le N$ and $\sigma^{2}=1$. In Fig. \ref{Rank-N} we compare our theory with the Monte-Carlo simulations for $N=1024$ dimensional matrices. In the main figure, we show results for $\mu=1$ and $3$ while for $\mu=0$ we plot only the Mar\v cenko Pastur density. As can be seen from this figure that the density tends to attain a uniform shape as $\mu$ is increased. This is closely predicted by the theory. In two insets, (a) and (b), we show result for $\mu=10$. As shown in (a) our theory gives reasonable account of the data through out the support for the density. In (b), we notice oscillations for $\lambda<1$ which is almost consistent with the theory.

\section{Pastur equation for nc-CWE}\label{nc-CWE}
Having specialized in CWE and nc-WE cases we now consider $\xi\ne\mathbf{1}_{N}$ and $\zeta\ne0$ in Eq. (\ref{g11},\ref{g22}). We first note that Eq. (\ref{g22}) can be written as
\begin{eqnarray}
\overline{g}_{22}&=&\frac{1-\kappa}{u-\sigma^{2}\kappa\overline{g}_{11;\xi}}+ \kappa 
\Big
\langle
\Big[
(u-\sigma^{2}\kappa\overline{g}_{11;\xi})\mathbf{1}_{N}
\nonumber\\
&-&\zeta (u-\sigma^{2}\xi \overline{g}_{22})^{-1}
\Big
]^{-1}
\Big
\rangle_{N}.
\end{eqnarray}
Using this and Eq. (\ref{g11}) one finds the relation (\ref{g11g22}) and consequently the relation (\ref{gzgu}). Next, exploiting relations (\ref{g11g22}) and (\ref{gzgu}) with $u^{2}=z$ in Eq. (\ref{g11}) we obtain a coupled Pastur equation for nc-CWE:
\begin{equation}\label{Pastur-nc-CWE}
\mathbf{\overline{g}}_{\mathbb{W};L}(z)=
\llan L
\frac{1}
{z\mathbf{1}_{N}-\alpha_{1}(z,\mathbf{\overline{g}}_{\mathbb{W}}(z))\,\xi-\alpha_{2}(\mathbf{\overline{g}}_{\mathbb{W};\xi}(z))\zeta}
\rran_{N},
\end{equation}
where $L$ is an arbitrary $N\times N$ matrix and
\begin{eqnarray}\label{alphabeta}
\alpha_{1}(z,\mathbf{\overline{g}}_{\mathbb{W}}(z))&=&\sigma^{2}(1-\kappa+\kappa z\mathbf{\overline{g}}_{\mathbb{W}}(z)),\nonumber\\
\alpha_{2}(\mathbf{\overline{g}}_{\mathbb{W};\xi}(z))&=&[1-\sigma^{2}\kappa\mathbf{\overline{g}}_{\mathbb{W};\xi}(z)]^{-1}.
\end{eqnarray}
Choices $L=\mathbf{1}_{N}$ and $L=\xi$ yield respectively $\mathbf{\overline{g}}_{\mathbb{W}}(z)$ and $\mathbf{\overline{g}}_{\mathbb{W};\xi}(z)$ and thus complete the result. It is easy to see that results (\ref{Pastur-CWE}) and (\ref{Pastur-NcW}) are immediate from the result  (\ref{Pastur-nc-CWE}), for the choices $L=\mathbf{1}_{N}$ and $\zeta=0$, and $L=\mathbf{1}_{N}$ and $\xi=\mathbf{1}_{N}$, respectively in (\ref{Pastur-nc-CWE}). 

\subsection{Separation of eigenvalues}
It is also important to note that in general Eq. (\ref{Pastur-nc-CWE}) can not be simplified to the eigenvalues of $\xi$ and $\zeta$ unless they commute with each other. Therefore, unlike the Pastur equation, it is difficult to extend the results (\ref{Ebar-CWE}, \ref{Sprt-NCW}) to the nc-CWE case. %It is also important to note that in general Eq. (\ref{Pastur-nc-CWE}) can not be simplified to the eigenvalues of $\xi$ and $\zeta$ unless they commute with each other. 

We consider $\xi_{jk}=\delta_{jk}+(1-\delta_{jk})\mu^{2}_{0}$ and $B_{jk}=\mu\delta_{jk}$. Note that $\xi$ is a diagonal plus a rank-$1$ matrix and $\mathsf{B}$ is also a rank-$1$ matrix. In this case we can write Eq. (\ref{Pastur-nc-CWE}) as 
\begin{equation}
\mathbf{\overline{g}}_{\mathbb{W}}(z)
=\overline{\mathbf{g}^{(0)}_{\mathbb{W}}}(z)+\frac{1}{N} \frac{1}{z-\overline{\lambda}_{N}}.
\end{equation}
Here, since $(N-1)$ eigenvalues of $\zeta$ are identically zero, we have
\begin{equation}
\overline{\mathbf{g}^{(0)}_{\mathbb{W}}}(z)=
\llan\mathbb{Q}_{N}
\frac{1}
{z\mathbf{1}_{N}-\alpha_{1}(z,\mathbf{\overline{g}_{W}}(z))\,\xi}
\rran_{N},
\end{equation}
where $\mathbb{Q}_{k}$ corresponds to the $k'$th eigenstates of $\xi$ and
\begin{equation}\label{eqLB}
\overline{\lambda}_{N}=
\lambda^{(\xi)}_{N}\,\alpha_{1}(\overline{\lambda}_{N},\mathbf{\overline{g}}_{\mathbb{W}}(\overline{\lambda}_{N}))+
\lambda^{(\zeta)}_{N}\,\alpha_{2}(\mathbf{\overline{g}}_{\mathbb{W};\xi}(\overline{\lambda}_{N})).
\end{equation}
Next, we write
\begin{eqnarray}\label{gWxi}
\overline{\mathbf{g}_{\mathbb{W};\xi}}(z)&=&\overline{\mathbf{g}_{\mathbb{W};\xi}^{(0)}}(z)+O(N^{-1}),\nonumber\\
\overline{\mathbf{g}_{\mathbb{W};\xi}^{(0)}}(z)
&=&
\llan\mathbb{Q}_{N}\,\xi
\frac{1}
{z\mathbf{1}_{N}-\alpha_{1}(z,\mathbf{\overline{g}}_{\mathbb{W}}(z))\xi}
\rran_{N}.
\end{eqnarray}

We notice a relation between $\alpha_{1}(z,\overline{\mathbf{g}^{(0)}_{\mathbb{W}}}(z))$ and $\alpha_{2}(z,\overline{\mathbf{g}^{(0)}_{\mathbb{W};\xi}}(z))$:
\begin{equation}\label{iden-alphabeta}
\alpha_{1}(z,\overline{\mathbf{g}^{(0)}_{\mathbb{W}}}(z))=\sigma^{2}\alpha_{2}(\overline{\mathbf{g}^{(0)}_{\mathbb{W};\xi}}(z)).
\end{equation}
To obtain the above result we write $\overline{\mathbf{g}^{(0)}_{\mathbb{W}}}(z)$ in terms of $\overline{\mathbf{g}^{(0)}_{\mathbb{W};\xi}}(z)$ and then use Eq. (\ref{alphabeta}). This relation simplifies Eq. (\ref{eqLB}) as
\begin{equation}\label{LB-beta}
\overline{\lambda}_{N}=\alpha_{2}(\overline{\mathbf{g}^{(0)}_{\mathbb{W};\xi}}(z))\, [\sigma^{2}\lambda^{(\xi)}_{N}+\lambda^{(\zeta)}_{N}].
\end{equation}
Further, using the above equation in Eq. (\ref{gWxi}), we find
\begin{equation}\label{gxi-beta}
\overline{\mathbf{g}_{\mathbb{W};\xi}^{(0)}}(\overline{\lambda}_{N})=[\alpha_{2}(\mathbf{\overline{g}}_{\mathbb{W};\xi}(\overline{\lambda}_{N}))]^{-1}\Psi_{N},
\end{equation}
where 
\begin{equation}\label{Psi}
\Psi_{N}=\llan
\mathbb{Q}_{N}\, \xi
\frac{1}{\sigma^{2}(\lambda^{(\xi)}_{N}\mathbf{1}_{N}-\xi)+\lambda^{(\zeta)}_{N}\mathbf{1}_{N}}
\rran_{N}.
\end{equation}
Substituting Eq. (\ref{gxi-beta}) in the definition of $\alpha_{2}$, we find
\begin{equation}\label{beta-Psi}
\alpha_{2}(\mathbf{\overline{g}}_{\mathbb{W};\xi}(\overline{\lambda}_{N}))=1+\sigma^{2}\kappa \Psi_{N}.
\end{equation}
Finally, we use the above result in (\ref{LB-beta}) and obtain% $\overline{\lambda}_{N}$ in terms of the spectrum of $\xi$ and $\zeta$, as
\begin{equation}\label{Sprt-nc-CWE}
\overline{\lambda}_{N}=(1+\sigma^{2}\kappa \Psi_{N})(\sigma^{2}\lambda^{(\xi)}_{N}+\lambda^{(\zeta)}_{N}).
\end{equation}

This result can be generalized to the block-diagonal $\xi$ and $\zeta$ with dimensionally the same blocks where each block of $\xi$ is represented by equal-cross-correlation matrix while the corresponding $\zeta$ block is of the rank-$1$. For this setup one can generalize results (\ref{Psi}) and (\ref{Sprt-nc-CWE}) replacing the subscript $N$ by $k$ for the $k$'th separated eigenvalue.

Solving the above equation for equal-cross-correlation matrix $\xi$ and a rank-$1$ matrix $\zeta$ we obtain the ensemble averaged mean position for the separated eigenvalue as
\begin{equation}
\overline{\lambda}_{N}=\frac{(N\Delta^{2}+\sigma^{2})(N\Delta^{2}+\sigma^{2}\kappa)}{N\Delta^{2}},~\Delta^{2}=\mu_{0}^{2}\sigma^{2}+\mu^{2},
\end{equation}
where the above result is valid for $N\Delta^{2}>\sqrt{\kappa}$. This result is an interesting generalization of the corresponding results for the CWE and nc-WE where the bulk density is described by the Mar\v cenko Pastur density with a rescaled variance $\sigma^{2}(1-\mu_{0}^{2})$ as in Eq. (\ref{den-eqcor}).
  %(\ref{Ebar-CWE}, \ref{Sprt-NCW}) respectively for 
\begin{figure}
        \centering
               \includegraphics [width=0.5\textwidth]{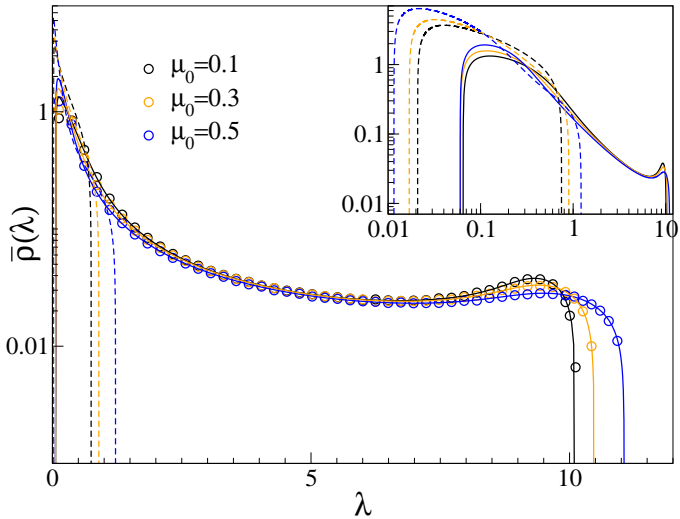}
                \caption{(Color online) The spectral density for the nc-CWOE where $\xi_{jk}=\delta_{jk}+(1-\delta_{jk})\mu_{0}^{(|j-k|)}$ and $\mathsf{B}_{j\nu}=\mu^{|j-\nu|}$ with  $\mu=0.5$, $\sigma^{2}=0.25$. In the main figure the density is shown on semi-log plot for $\mu_{0}=0.1,\,0.3$ and $0.5$ respective with {\it black, orange} and {\it blue} colors. Solids lines represent the theory (\ref{Pastur-nc-CWE}) for nc-CWE and dashes lines represent the corresponding $\mu=0$ cases. In the inset, theories for $\mu=0$ and $\mu=0.5$ are compared on log-log scale.}
\label{excr}
\end{figure}

\subsection{A non-trivial Example}
For non-trivial $\xi$ and $\zeta$ Eq. (\ref{Pastur-nc-CWE}) can be solved numerically. Thus one has to extend the algorithm for two equations of two variables, viz.
\begin{eqnarray}
f_{1}(\overline{\mathbf{g}^{(n)}_{\mathbb{W}}}(z), \overline{\mathbf{g}^{(n)}_{\mathbb{W};\xi}}(z))-\overline{\mathbf{g}^{(n)}_{\mathbb{W}}}(z)&=&0\nonumber\\
f_{2}((\overline{\mathbf{g}^{(n)}_{\mathbb{W}}}(z),\overline{\mathbf{g}^{(n)}_{\mathbb{W};\xi}}(z))-\overline{\mathbf{g}^{(n)}_{\mathbb{W};\xi}}(z)&=&0,
\end{eqnarray}
where $f_{1}(\overline{\mathbf{g}^{(n)}_{\mathbb{W}}}(z),\overline{\mathbf{g}^{(n)}_{\mathbb{W};\xi}}(z))$ and $f_{2}((\overline{\mathbf{g}^{(n)}_{\mathbb{W}}}(z),\overline{\mathbf{g}^{(n)}_{\mathbb{W};\xi}}(z))$ are the r.h.s of (\ref{Pastur-nc-CWE}) respectively with $L=\mathbf{1}_{N}$ and $L=\xi$. Next, we start with initial guesses $\overline{\mathbf{g}^{(0)}_{\mathbb{W}}}(z)$ and $\overline{\mathbf{g}^{(0)}_{\mathbb{W};\xi}}(z)$ for a given $z$ and use the Newton's method to obtain the solution in the machine precision. 

To illustrate the result we solve Eq. (\ref{Pastur-nc-CWE}) for $\xi_{jk}=\delta_{jk}+(1-\delta_{jk})\mu_{0}^{(|j-k|)}$, where $\mathsf{B}_{j\nu}=\mu^{|j-\nu|}$ with $\mu=0.5$ and $\mu_{0}$ is varied as $\mu_{0}=0.1,\,0.3$ and $0.5$. Also we choose $\sigma^{2}=0.25$ and $N=512$ with $T=2N$. The result is shown in Fig. \ref{excr} where open circles represent the histogram data obtained from the Monte-Carlo simulation of $\mathsf{C}$ and solid lines are obtained from the numerical solution of the theory (\ref{Pastur-nc-CWE}) where we have considered $N=512$. As shown in the figure, the theory reasonably explains numerical results. In this figure we also compare theory for $\mu=0.5$ for the corresponding CWOE ($\mu=0$). As can be seen in this figure, the nonzero mean not only changes the density profile but also shifts non-trivially the spectrum.  

\section{Summary and Discussions} \label{conclusion}
We have studied nc-CWE and obtained exact result for the spectral density at large matrix dimension. The derivation is formalized in two steps, viz. first we obtain the loop equation for $\mathsf{X}$, which eigenvalues are closely related with those of $\mathbb{W}$, and secondly we derive the Pastur equation for $\mathbb{W}$ from the loop equation. With this formalism we have derived the Pastur equation for CWE, nc-WE and for the nc-CWE. For all the three cases we have exploited a linear relation between the averaged quantities $u\overline{g}_{11}$ and $u\overline{g}_{22}$. We notice that in the first two cases the Pastur equation depends on the eigenvalues of positive definite symmetric matrices, $\xi$ and $\zeta=\mathsf{BB}^{\dagger}/T$. We have shown that in general, unlike CWE and nc-WE, the spectral density for nc-CWE does not depend simply on the spectra of $\xi$ and $\zeta$ rather more intricately on the matrices.

From the Pastur equation, we have worked out the ensemble-averaged mean position of the separated eigenvalues for the nc-WE. For CWE this has been worked out in Ref. \cite{Baik:2005, Forrester, vp2010}. Following Ref. \cite{vp2010}, we have given the result for a general $\zeta$ in the nc-WE case. In nc-CWE case the the Pastur equation is more complicated. However, we have been able to worked out the ensemble-averaged mean position of the separated eigenvalues for some especial cases of nc-CWE. As for the CWE and nc-CWE, for more general cases we have used the Newton's method to solve the Pastur equation numerically. We have supplemented our theoretical result with numerics with some non-trivial examples.

Finally, it would be interesting to extend this generalization for the Wishart model of nonsymmetric correlation matrices those dealt in Refs. \cite{vin2013, vinLuis:2014}. Another important extension of this work is related to short time series often encountered in the correlation analysis of multivariate complex systems. For short time series, $N>T$ resulting a correlation matrix which is singular with significantly many zero eigenvalues. In Ref. \cite{GuhrKaebler,GuhrShaefer} the power map method is proposed and used recently in \cite{VBPS:2014} as tool to get rid of this degeneracy. This method results an spectrum emerging from the zero eigenvalues when the exponent is very close to $1$. It has been shown in Ref. \cite{vrt:2013} that the so emerging spectrum is very sensitive to correlations and we believe that the study of the emerging spectra corresponding nc-CWE is very important. 

\section{Acknowledgments}
The author is grateful to Thomas H. Seligman and Luis Benet for useful discussions. Financial support from CONACyT through Project No. 154586 and No. PAPIIT UNAM RR 113311 is acknowledged. The author was supported by DGAPA/UNAM as a postdoctoral fellow.

\end{document}